\documentclass[conference]{IEEEtran}

\usepackage{amsmath,amssymb,amsfonts,bm}
\usepackage{graphicx}
\usepackage{booktabs}
\usepackage{cite}
\usepackage{mathtools}
\usepackage[caption=false,font=footnotesize]{subfig}

\graphicspath{{../figures/}}

\begin{document}

\title{Privacy-Preserving Federated Radio Map Learning\\for Wireless Digital Twins\\via Adaptive Noise Allocation}

\author{\IEEEauthorblockN{Jijia Tian\IEEEauthorrefmark{1}\IEEEauthorrefmark{3},
        Hao Wang\IEEEauthorrefmark{1}\IEEEauthorrefmark{3},
        Mu Jia\IEEEauthorrefmark{1}\IEEEauthorrefmark{2}\IEEEauthorrefmark{3},
        Yi Wang\IEEEauthorrefmark{1}\IEEEauthorrefmark{3},
		Junting Chen\IEEEauthorrefmark{1}\IEEEauthorrefmark{3} and
		Pooi-Yuen Kam\IEEEauthorrefmark{2}\IEEEauthorrefmark{3}}
	
	\IEEEauthorblockA{\IEEEauthorrefmark{1}School of Science and Engineering,
		\IEEEauthorrefmark{2}School of Artificial Intelligence,\\
		\IEEEauthorrefmark{3}Shenzhen Future Network of Intelligence Institute (FNii-Shenzhen),\\
		The Chinese University of Hong Kong (Shenzhen), Guangdong 518172, China}
}

\maketitle

\begin{abstract}
Radio maps provide a foundational data layer for wireless digital twins, and federated learning offers a natural framework for their distributed construction without centralizing raw radio environment data. However, the exchanged client model updates may still leak transmitter-location information, even when the underlying measurement data are never shared. Existing noise-based privacy defenses inject perturbation either uniformly across all uploaded coordinates or according to a fixed static rule, thereby ignoring the architecture-specific structure of this leakage. This paper proposes a budget-constrained adaptive noise allocation mechanism that redistributes a fixed perturbation budget across transmitter-sensitive upload groups identified from the two-stage RadioUNet architecture. The proposed method uses low-dimensional upload statistics to dynamically adjust group-wise noise scales and is integrated locally before client upload transmission. We evaluate the framework on a federated radio map learning task under a unified noise multiplier, comparing it against uniform and structure-aware baselines using reconstruction mean squared error and transmitter localization error as metrics. Results show that adaptive allocation achieves the strongest privacy protection while maintaining the best reconstruction quality among all noise-based defenses under a matched perturbation budget.
\end{abstract}

\section{Introduction}
\label{sec:intro}

Radio maps have emerged as a key enabler for environment-aware wireless system design. By providing spatially continuous representations of signal propagation, radio maps support channel knowledge acquisition~\cite{ZenCheXu:J24}, data-driven radio map estimation~\cite{LevYapKut:J21, JiaSunChe:C26}, low-altitude sensing, localization, and communication~\cite{SunJiaYu:C25, liZhaJia:J25, LiaJiaZha:J26, TiaCheKam:J26}, and communication optimization in dynamic vehicular environments~\cite{Wan:J25}. In practical deployments, the measurement data required for radio map construction are inherently distributed across geographically separated edge nodes, base stations, or user devices. Federated learning~\cite{McMaHam:C17} provides an attractive training paradigm by enabling collaborative model training without centralizing raw data. Recent work has demonstrated the viability of federated radio map construction with differential privacy guarantees~\cite{TiaCheChe:C24, TiaCheChe:J25}. Nevertheless, avoiding raw data sharing does not automatically eliminate privacy leakage from the exchanged model updates.

Recent studies on gradient leakage~\cite{LiZhaLiu:C22} have demonstrated that model updates transmitted during federated training can reveal sensitive information about the local training data. In federated radio map learning, this risk takes a distinctive form: because the transmitter raster is explicitly encoded as an input channel to the RadioUNet architecture~\cite{LevYapKut:J21}, the uploaded client updates carry structured transmitter-dependent fingerprints that can be exploited for localization. Existing differentially private federated learning frameworks~\cite{WeiLiDin:J20} inject noise uniformly across all update coordinates, treating the leakage as homogeneous. Recent work on location-safe federated radio map construction~\cite{TiaCheChe:J25} has introduced geometry-aligned differential privacy that accounts for the spatial structure of location information, yet the perturbation allocation remains fixed across upload coordinates rather than adapting to the architecture-specific leakage pattern during training. In short, no existing mechanism dynamically distributes a fixed perturbation budget across upload groups according to their evolving architecture-specific leakage sensitivity.

To address this gap, we develop a budget-constrained adaptive noise allocation mechanism for privacy-preserving federated radio map learning, designed for integration into digital twin pipelines where distributed radio map construction must satisfy location-privacy requirements. The main contributions are summarized as follows:
\begin{itemize}
\item We formulate upload-based transmitter-location leakage as a privacy threat in federated radio map learning for wireless digital twin applications and define a corresponding budget-constrained privacy-utility optimization problem.
\item We propose a structure-aware adaptive noise allocator that dynamically distributes a fixed perturbation budget across transmitter-sensitive upload groups according to the current upload state.
\item We establish an evaluation protocol for federated RadioUNet under upload-based localization attacks and demonstrate that adaptive allocation outperforms both uniform and static structure-aware baselines under a matched perturbation budget.
\end{itemize}

\section{System Model}
\label{sec:system}

We consider a federated radio map learning system in which multiple clients collaboratively train a shared RadioUNet predictor without exchanging raw measurement data. Although this architecture avoids centralizing sensitive radio environment observations, the model updates uploaded by each client may still encode transmitter-location information that a passive observer can exploit. To formalize this tension, we first describe the federated learning architecture and the two-stage RadioUNet backbone (Section~II-A), then define the upload-based transmitter localization threat (Section~II-B), and finally cast the resulting privacy-utility tradeoff as a budget-constrained optimization problem (Section~II-C).

\subsection{Federated Radio Map Learning Model}
Consider a set of clients $\mathcal{K} = \{1,\ldots,K\}$. Client $k \in \mathcal{K}$ owns a local map subset $\mathcal{M}_k$, and each map contains multiple transmitter realizations indexed by $t$. For each triplet $(k,m,t)$ with $m \in \mathcal{M}_k$, the input tensor is denoted by $\mathbf{x}_{k,m,t} \in \mathbb{R}^{N_c \times H \times W}$ and the target radio map is denoted by $\mathbf{y}_{k,m,t} \in \mathbb{R}^{1 \times H \times W}$, where $N_c$ is the number of input channels and $H \times W$ is the spatial resolution. In the RadioUNet setting here, $\mathbf{x}_{k,m,t}$ contains the building map and the transmitter raster, while $\mathbf{y}_{k,m,t}$ is the corresponding pathloss map.

The reconstruction network follows the two-stage RadioUNet architecture. The first stage produces a coarse estimate,
\begin{equation}
\hat{\mathbf{y}}_{k,m,t}^{(1)} = f_{\boldsymbol{\theta}_1}\!\left(\mathbf{x}_{k,m,t}\right),
\label{eq:first_stage}
\end{equation}
and the second stage refines this estimate by using both the original input and the first-stage output,
\begin{equation}
\hat{\mathbf{y}}_{k,m,t} = f_{\boldsymbol{\theta}_2}\!\left(\mathbf{x}_{k,m,t}, \hat{\mathbf{y}}_{k,m,t}^{(1)}\right).
\label{eq:second_stage}
\end{equation}
Following the standard RadioUNet training protocol~\cite{LevYapKut:J21}, the first stage is trained first, and the second stage is then optimized using the first-stage prediction as its input context.

For a given client model parameter vector $\boldsymbol{\theta}$, the per-sample reconstruction loss is
\begin{equation}
\ell\!\left(\boldsymbol{\theta}; \mathbf{x}_{k,m,t}, \mathbf{y}_{k,m,t}\right)
= \left\| \hat{\mathbf{y}}_{k,m,t} - \mathbf{y}_{k,m,t} \right\|_2^2,
\label{eq:sample_loss}
\end{equation}
where the prediction in \eqref{eq:sample_loss} refers to the stage currently being optimized. The empirical loss at client $k$ is
\begin{equation}
F_k(\boldsymbol{\theta})
= \frac{1}{|\mathcal{D}_k|}
\sum_{(\mathbf{x},\mathbf{y}) \in \mathcal{D}_k}
\ell(\boldsymbol{\theta}; \mathbf{x}, \mathbf{y}),
\label{eq:client_loss}
\end{equation}
where $\mathcal{D}_k$ denotes the local dataset of client $k$. The global learning target is the weighted sum
\begin{equation}
F(\boldsymbol{\theta}) = \sum_{k \in \mathcal{K}} p_k F_k(\boldsymbol{\theta}),
\label{eq:global_loss}
\end{equation}
where $p_k$ is proportional to the number of local samples at client $k$.

At communication round $r$, the server broadcasts the current global model $\boldsymbol{\theta}_r$ to the participating clients. Each selected client performs local training and then uploads a model update. Since the focus of this paper is privacy protection rather than the design of a new federated optimizer, the aggregation rule is kept general at this stage and will be treated as the backend update operator in Section~III.

\subsection{Upload Leakage Model}

The privacy threat considered in this work arises from the uploaded client updates rather than from the raw input data. Let $\boldsymbol{\theta}_{k,r}^{\mathrm{loc}}$ denote the local model obtained by client $k$ at round $r$ after its local training stage. The uploaded client update is
\begin{equation}
\Delta \boldsymbol{\theta}_{k,r}
= \boldsymbol{\theta}_{k,r}^{\mathrm{loc}} - \boldsymbol{\theta}_r.
\label{eq:client_delta}
\end{equation}
The observer in our threat model has access to the client uploads and attempts to infer the transmitter coordinate associated with the local sample that generated the update pattern.

The key observation is that the leakage is structured rather than uniformly distributed over all uploaded parameters. Because the transmitter raster is explicitly provided as an input channel to the RadioUNet architecture, the uploaded gradients or model deltas are expected to be more informative in transmitter-coupled components than in the rest of the parameter space. Accordingly, the attacker first extracts an upload fingerprint from a short local update trajectory and then regresses the transmitter coordinate from that fingerprint. Denote by $\mathbf{c}_{k,m,t} \in \mathbb{R}^2$ the true transmitter coordinate, by $\phi(\cdot)$ the fingerprint extraction operator, and by $h(\cdot)$ the localization attacker. The attack model is written as
\begin{equation}
\hat{\mathbf{c}}_{k,m,t}
= h\!\left(\phi\!\left(\Delta \boldsymbol{\theta}_{k,m,t}^{(1:S)}\right)\right),
\label{eq:attacker_model}
\end{equation}
where $\Delta \boldsymbol{\theta}_{k,m,t}^{(1:S)}$ denotes the sequence of uploads associated with $S$ local update steps.

\begin{figure}[t]
\centering
\includegraphics[width=0.95\columnwidth]{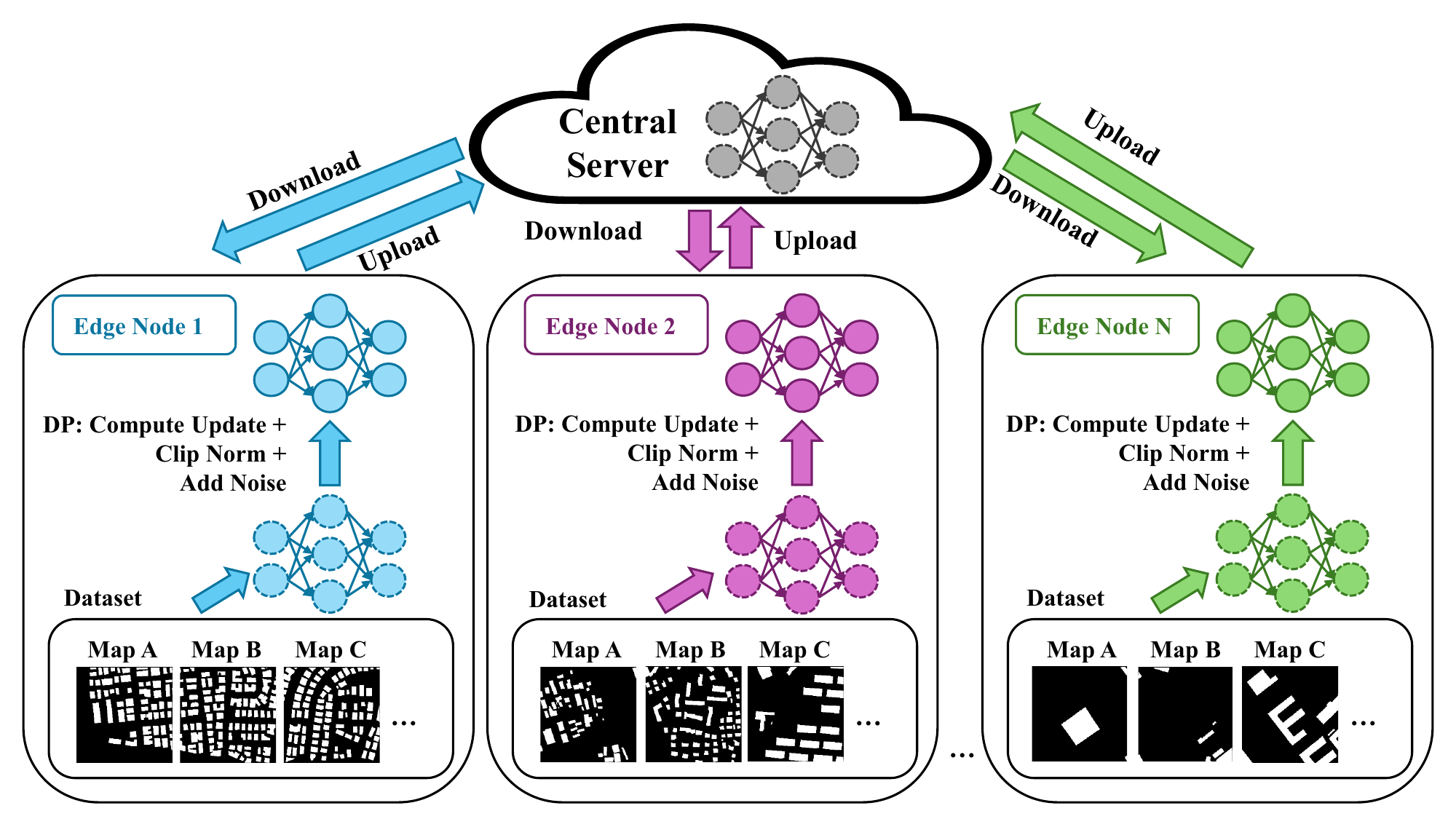}
\caption{System overview. Each client holds local radio map data (building map, transmitter raster, and pathloss target) and trains a two-stage RadioUNet. The raw model update is passed through a client-side defense module before transmission to the server. A passive attacker observes the raw upload and attempts to infer the transmitter location.}
\label{fig:system}
\end{figure}

The privacy metric is the localization error
\begin{equation}
e_{\mathrm{loc}}
= \left\| \hat{\mathbf{c}}_{k,m,t} - \mathbf{c}_{k,m,t} \right\|_2.
\label{eq:loc_error}
\end{equation}
This section only defines the threat model and the privacy target. Quantitative attack results are deferred to Section~IV. Fig.~\ref{fig:system} illustrates the overall system architecture including the leakage path and defense insertion point.

\subsection{Problem Formulation}

The objective of this work is not to redesign federated radio map learning itself, but to perturb the uploaded client update before transmission so that localization becomes more difficult while reconstruction performance remains acceptable. To make the tradeoff interpretable, the perturbation is constrained by a fixed total budget per upload.

Let the uploaded parameter space be partitioned into $G$ groups, and let $E_{k,r,g}$ denote the perturbation energy assigned to group $g$ for client $k$ at round $r$. The fixed-budget constraint is
\begin{equation}
\sum_{g=1}^{G} E_{k,r,g} = B, \qquad E_{k,r,g} \ge 0,
\label{eq:budget_constraint}
\end{equation}
where $B$ is the total perturbation budget. The design goal is then to find an allocation policy $\pi$ that improves privacy while controlling utility degradation. A compact form of the resulting optimization problem is
\begin{equation}
\min_{\pi}
\mathcal{L}_{\mathrm{task}}(\pi)
- \lambda \, \mathcal{L}_{\mathrm{priv}}(\pi)
\quad
\text{s.t.}
\quad
\sum_{g=1}^{G} E_{k,r,g} = B,
\label{eq:problem_formulation}
\end{equation}
where $\mathcal{L}_{\mathrm{task}}$ captures reconstruction quality, $\mathcal{L}_{\mathrm{priv}}$ captures attacker success, and $\lambda > 0$ controls the privacy-utility tradeoff. In this paper, the proposed policy is designed as a structure-aware adaptive budget allocator. We emphasize that the goal is a budgeted adaptive privacy defense; the paper does not claim a full end-to-end differential privacy accountant for the adaptive mechanism.

\section{Proposed Adaptive Budgeted Privacy Defense}

\subsection{Adaptive Noise Allocation}

Since transmitter-related leakage is highly uneven across the uploaded parameter space whereas the perturbation budget is fixed, uniform perturbation is generally inefficient because it spends part of the budget on components that carry little location information. We therefore partition the uploaded parameter space into a set of transmitter-sensitive groups and one residual group. Let $\mathcal{G} = \{1,\ldots,G\}$ denote the group index set, and let $\mathbf{M}_g \in \{0,1\}^{d}$ denote the binary selector mask for group $g$. The uploaded client update can then be written as
\begin{equation}
\Delta \boldsymbol{\theta}_{k,r}
= \sum_{g=1}^{G} \mathbf{M}_g \odot \Delta \boldsymbol{\theta}_{k,r},
\label{eq:group_decomposition}
\end{equation}
where $\odot$ denotes the Hadamard product. In the current implementation, the transmitter-sensitive groups correspond to the parameter subsets directly coupled to the transmitter channel in the first and second RadioUNet stages, while the residual group contains the remaining parameters.

Instead of feeding the full client upload into the allocator, we extract a compact group-wise summary vector. Let $\Psi(\cdot)$ denote the statistics extractor that collects, for each group, the masked $\ell_2$ norm, mean magnitude, standard deviation, and group size ratio, together with the global upload norm, the normalized round index, and the phase indicator. The resulting low-dimensional descriptor is
\begin{equation}
\mathbf{s}_{k,r}
= \Psi\!\left(\Delta \boldsymbol{\theta}_{k,r}\right).
\label{eq:allocator_input}
\end{equation}
This design keeps the control mechanism low-dimensional and interpretable, while still allowing the perturbation pattern to adapt to the current upload state.

The adaptive allocator is parameterized by a neural network $f_{\boldsymbol{\eta}}(\cdot)$ that maps the summary vector in \eqref{eq:allocator_input} to allocation logits,
\begin{equation}
\mathbf{a}_{k,r} = f_{\boldsymbol{\eta}}\!\left(\mathbf{s}_{k,r}\right).
\label{eq:allocator_logits}
\end{equation}
The normalized budget weights are obtained by
\begin{equation}
\mathbf{w}_{k,r} = \mathrm{softmax}\!\left(\mathbf{a}_{k,r}\right),
\label{eq:allocator_weights}
\end{equation}
and the group-wise energy allocation is
\begin{equation}
E_{k,r,g} = B \, w_{k,r,g}, \qquad g \in \mathcal{G}.
\label{eq:energy_allocation}
\end{equation}
which guarantees that the total perturbation budget remains fixed while the distribution across groups is adaptive.

To match the total second-moment budget of a uniform Gaussian baseline, let $\nu$ denote the noise multiplier and define the baseline noise scale as $\sigma_0 = C\nu$, where $C > 0$ is the $\ell_2$ clipping threshold introduced below. The total perturbation budget is then $B = d\,\sigma_0^2$, where $d = \sum_{g \in \mathcal{G}} d_g$ is the total number of upload coordinates and $d_g$ denotes the number of parameters in group $g$. The group-wise Gaussian perturbation scale is defined by
\begin{equation}
\sigma_{k,r,g}
= \sqrt{\frac{E_{k,r,g}}{d_g}}.
\label{eq:group_sigma}
\end{equation}
Before noise injection, the client update is clipped with threshold $C$,
\begin{equation}
\overline{\Delta \boldsymbol{\theta}}_{k,r}
= \mathrm{Clip}_C\!\left(\Delta \boldsymbol{\theta}_{k,r}\right),
\label{eq:clipped_delta}
\end{equation}
and the privatized upload becomes
\begin{equation}
\widetilde{\Delta \boldsymbol{\theta}}_{k,r}
= \overline{\Delta \boldsymbol{\theta}}_{k,r}
+ \sum_{g=1}^{G} \mathbf{M}_g \odot \mathbf{n}_{k,r,g},
\label{eq:privatized_upload}
\end{equation}
where
\begin{equation}
\mathbf{n}_{k,r,g}
\sim \mathcal{N}\!\left(\mathbf{0}, \sigma_{k,r,g}^2 \mathbf{I}\right).
\label{eq:group_noise}
\end{equation}
The perturbation budget is therefore fixed at the upload level, while the distribution of this budget across sensitive groups is learned adaptively.

\subsection{Training Objective and Aggregation-Aware Interpretation}

The adaptive allocator is trained through a proxy attack model. The reason is that the final attacker used for evaluation can be non-differentiable or external to the training loop, whereas the allocator itself requires a differentiable supervision signal. Let $\varphi(\cdot)$ denote the fingerprint extraction operator applied to a privatized upload, and let $q_{\boldsymbol{\psi}}(\cdot)$ denote the proxy attacker that predicts the transmitter coordinate from that fingerprint. For a proxy training batch of $N$ upload traces indexed by $i$, the proxy loss is
\begin{equation}
\mathcal{L}_{\mathrm{proxy}}
= \frac{1}{N}
\sum_{i=1}^{N}
\left\|
q_{\boldsymbol{\psi}}\!\left(\varphi\!\left(\widetilde{\Delta \boldsymbol{\theta}}_i\right)\right)
- \mathbf{c}_i
\right\|_2^2.
\label{eq:proxy_loss}
\end{equation}
The proxy attacker is trained to minimize \eqref{eq:proxy_loss}, while the allocator is trained to make the attack more difficult under the task-quality constraint.

To balance privacy and utility, we use the surrogate allocator objective
\begin{equation}
\mathcal{L}_{\mathrm{alloc}}
= \mathcal{L}_{\mathrm{task}}
- \lambda_p \mathcal{L}_{\mathrm{proxy}}
- \lambda_h \mathcal{H}\!\left(\mathbf{w}_{k,r}\right),
\label{eq:allocator_loss}
\end{equation}
where $\mathcal{L}_{\mathrm{task}}$ denotes the reconstruction loss, $\mathcal{H}(\cdot)$ is the entropy of the allocation vector, and $\lambda_p,\lambda_h > 0$ are balancing coefficients. The second term encourages the allocator to reduce attacker accuracy, while the entropy regularizer mitigates degenerate solutions that collapse the budget into a single group. In implementation, the proxy attacker and the allocator are updated in alternating steps within each federated round.

The effect of the injected Gaussian noise should be interpreted carefully after server aggregation. For clarity, consider the averaged client update before the server optimizer is applied. If $K_r$ clients participate at round $r$, the aggregated privatized update is
\begin{equation}
\overline{\widetilde{\Delta \boldsymbol{\theta}}}_{r}
= \frac{1}{K_r}
\sum_{k=1}^{K_r}
\left(
\Delta \boldsymbol{\theta}_{k,r}
+ \mathbf{n}_{k,r}
\right)
= \overline{\Delta \boldsymbol{\theta}}_{r}
+ \overline{\mathbf{n}}_{r},
\label{eq:aggregated_update}
\end{equation}
where $\mathbf{n}_{k,r}$ denotes the total injected perturbation in \eqref{eq:privatized_upload}. Since the client perturbations are zero-mean, the aggregated perturbation also satisfies
\begin{equation}
\mathbb{E}\!\left[\overline{\mathbf{n}}_{r}\right] = \mathbf{0}.
\label{eq:agg_noise_mean}
\end{equation}
If the client perturbations are independent and have matched variance, then
\begin{equation}
\mathrm{Var}\!\left(\overline{\mathbf{n}}_{r}\right)
= \frac{1}{K_r^2}
\sum_{k=1}^{K_r}
\mathrm{Var}\!\left(\mathbf{n}_{k,r}\right),
\label{eq:agg_noise_var}
\end{equation}
which implies an effective standard-deviation scaling proportional to
\begin{equation}
\sigma_{\mathrm{eff}} \propto \frac{1}{\sqrt{K_r}}.
\label{eq:agg_noise_scale}
\end{equation}
Hence, aggregation attenuates the perturbation variance in the global model but does not remove privacy protection, because the attacker in our threat model observes individual client uploads before aggregation.

\section{Experiments and Results}
\label{sec:exp}

\subsection{Experimental Setup}
\label{subsec:setup}

The experiments are conducted on a dataset containing $700$ urban radio maps~\cite{TegRom:J21, LevYapKut:J21}. A map-level split is applied before client partitioning, and the reported results are averaged over multiple random seeds. The training process consists of $50$ communication rounds. In each round, $6$ clients are randomly selected from $14$ participating clients, and each selected client performs one local epoch with a batch size of~$64$. The local and server learning rates are identical across all schemes.

We compare six methods. \textit{No~Defense} uses the full two-stage federated training protocol without any privacy mechanism and serves as the task fidelity upper bound. \textit{FedSGD} uses vanilla single-step federated SGD without the enhanced training protocol, providing a weaker but simpler reference. \textit{Clip~Only} applies global $\ell_2$ norm clipping on the client upload without additional noise injection, isolating the effect of clipping from stochastic perturbation. \textit{Uniform~Noise} clips the client update and then injects Gaussian noise with identical strength into all upload coordinates. \textit{Directed~Uniform} applies uniform Gaussian perturbation only to the parameter subspace most strongly coupled with transmitter-location information, while leaving the remaining parameters unperturbed. \textit{Proposed} is the adaptive noise allocation method of this paper; it operates under the same total perturbation budget but dynamically redistributes the noise energy across sensitive parameter groups according to the leakage structure revealed during training. To ensure comparability, all explicit noise-based defenses use a unified noise multiplier of~$\nu = 3$. Reconstruction quality is measured by the best validation mean squared error~(MSE), where lower values indicate better task fidelity. Privacy protection is quantified by the root mean squared localization error of a proxy attacker in meters, where higher values indicate stronger privacy.

\subsection{Results and Discussion}
\label{subsec:results}

\begin{table}[t]
\caption{Privacy-utility comparison at noise multiplier $\nu = 3$. Lower MSE~(dB) indicates better reconstruction; higher RMSE indicates stronger privacy.}
\label{tab:main}
\centering
\begin{tabular}{lcc}
\toprule
Scheme & Val MSE~(dB)\,$\downarrow$ & Privacy RMSE~(m)\,$\uparrow$ \\
\midrule
No Defense       & $-16.16$ & 24.86 \\
Clip Only        & $-16.00$ & 26.85 \\
FedSGD           & $-15.26$ & 29.16 \\
Uniform Noise    & $-14.43$ & 36.78 \\
Directed Uniform & $-15.66$ & 42.26 \\
Proposed         & $\mathbf{-15.87}$ & $\mathbf{45.63}$ \\
\bottomrule
\end{tabular}
\end{table}

\begin{figure}[t]
\centering
\subfloat[Reconstruction MSE (dB) over rounds.]{\includegraphics[width=0.95\columnwidth]{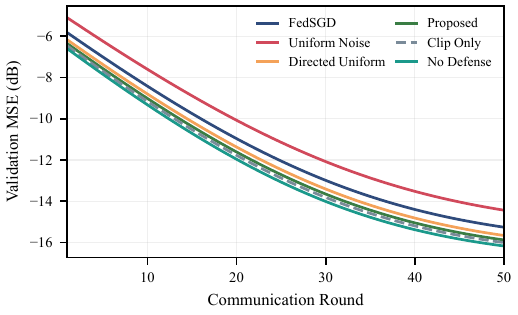}\label{fig:recon}}\\
\subfloat[Privacy proxy RMSE (m) over rounds.]{\includegraphics[width=0.95\columnwidth]{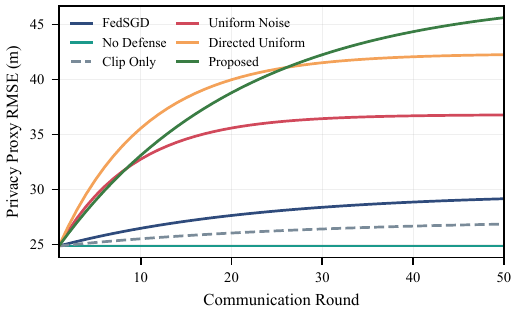}\label{fig:privacy}}
\caption{Evolution of reconstruction quality and privacy protection over $50$ communication rounds under noise multiplier $\nu = 3$.}
\label{fig:results}
\end{figure}

Table~\ref{tab:main} and Fig.~\ref{fig:results} present the main experimental results. Fig.~\ref{fig:recon} shows the reconstruction error in dB scale over $50$ communication rounds. All schemes converge stably, confirming that the federated training protocol remains effective under a limited communication budget. The No~Defense reference and Clip~Only achieve the lowest reconstruction error, indicating that global norm clipping alone has only a marginal impact on task fidelity. Uniform~Noise stays in the highest error regime throughout training, implying that indiscriminate perturbation across all coordinates degrades both location-sensitive cues and task-relevant representations. Both Directed~Uniform and the proposed method clearly outperform Uniform~Noise, and among all noise-based defenses, the proposed method achieves the lowest final reconstruction error with $-15.87$~dB, further improving upon Directed~Uniform~ with $-15.66$~dB in the later training stages.

Fig.~\ref{fig:privacy} reports the privacy protection over communication rounds. At round~$1$, all schemes start from nearly the same attacker localization error of approximately $24.86$~m. As training proceeds, No~Defense and Clip~Only exhibit only limited increases in attacker error. All three noise-based defenses provide a much more pronounced privacy gain, although their evolution patterns differ markedly. Uniform~Noise and Directed~Uniform rise rapidly in the early stage but gradually flatten, indicating that fixed perturbation rules offer diminishing marginal privacy benefit once the model adapts. The proposed method, by contrast, exhibits a persistent upward trajectory and reaches the highest final localization error of~$45.63$~m, compared to~$42.26$~m for Directed~Uniform and~$36.78$~m for Uniform~Noise.

These observations reveal two key insights. First, where noise is injected matters more than how much noise is injected. Directed~Uniform and the proposed method both outperform Uniform~Noise despite sharing the same total perturbation budget, confirming that transmitter-related leakage is concentrated in a small subset of the uploaded parameter space. Spending the budget on coordinates that carry little location information degrades task fidelity without proportional privacy benefit.

Second, dynamic reallocation provides gains beyond static structure-aware injection. While Directed~Uniform achieves strong privacy by concentrating noise on a predefined sensitive subspace, its fixed allocation rule cannot adapt to the evolving training state. The proposed method overcomes this limitation by continuously adjusting the budget distribution, which explains both its persistent privacy improvement over later rounds and its superior final reconstruction quality.

\section{Conclusion}
\label{sec:conclusion}

This paper studied upload-based transmitter-location leakage in federated radio map learning, a setting motivated by the growing need for privacy-preserving radio environment construction in wireless digital twin systems. We proposed a budget-constrained adaptive noise allocation framework that partitions the uploaded parameter space into architecture-aware transmitter-sensitive groups and dynamically redistributes a fixed perturbation budget across these groups using low-dimensional upload statistics. Experimental results confirm that the proposed method achieves the best privacy-utility tradeoff among all compared defenses under a matched perturbation budget. Future directions include robustness evaluation against diverse attacker models and formal differential privacy accounting.

\bibliographystyle{IEEEtran}
\bibliography{refs}

\end{document}